**Atomic-scale structure of the SrTiO$_3$_(001)-c(6x2) reconstruction: Experiments and first-principles calculations**

C. H. Lanier[1,2], A. van de Walle[3,*], N. Erdman[4,*], E. Landree[5,*], O. Warschkow[6,**], A. Kazimirov[7,***], K. R. Poeppelmeier[2,8], J. Zegenhagen[9,10], M. Asta[11,*] and L. D. Marks[1,2]

[1]Department of Materials Science and Engineering, Northwestern University, Evanston, IL 60208, USA

[2]Institute for Catalysis in Energy Processes, Northwestern University, Evanston, IL 60208, USA

[3]Engineering & Applied Science Division, California Institute of Technology, Pasadena, CA 91125, USA

[4]JEOL USA, Inc., 11 Dearborn Rd, Peabody, MA 01960

[5]RAND Corporation, Arlington, Virginia 22202

[6]School of Physics, The University of Sydney, Sydney, NSW 2026, Australia

[7]Cornell High Energy Synchrotron Source, Ithaca, NY 14953, USA

[8]Department of Chemistry, Northwestern University, Evanston, IL 60208, USA

[9]Max-Planck Institut für Festkörperforschung, Heisenbergstr. 1, D-70569 Stuttgart, Germany

[10]European Synchrotron Radiation Facility ESRF, BP 220, F-39043, Grenoble, France

[11]Chemical Engineering and Materials Science Department, University of California at Davis, Davis, CA 95616

*Work was performed while at Northwestern University, Department of Materials Science and Engineering

**Work was performed while at Northwestern University, Department of Physics and Astronomy

***Work was performed while at Max-Planck-Institute für Festkörperforschung

**ABSTRACT**

The c(6x2) is a reconstruction of the SrTiO$_3$(001) surface that is formed between 1050-1100$^o$C in oxidizing annealing conditions. This work proposes a model for the atomic structure for the c(6x2) obtained through a combination of results from transmission electron diffraction, surface x-ray diffraction, direct methods analysis, computational combinational screening, and density functional theory. As it is formed at high temperatures, the surface is complex and can be described as a short-range ordered phase featuring microscopic domains composed of four main structural motifs. Additionally, non-periodic TiO$_2$ units are present on the surface. Simulated scanning tunneling microscopy images based on the electronic structure calculations are consistent with experimental images.

# 1. INTRODUCTION

Strontium titanate ($SrTiO_3$) has received considerable attention over the last decade because of its numerous technological applications,[1] including use as a substrate for thin film growth[2] and as a candidate crystalline gate dielectric in silicon-based devices.[3, 4] Furthermore, the surface of $SrTiO_3$ plays an important role in surface reactions and catalysis.[5, 6] Many of these applications are governed by interfacial processes which motivates a continuing interest in the surfaces of $SrTiO_3$, but despite extensive research into the surface properties, there remain many important unanswered questions. Only recently the (2x1) and c(4x2) reconstructions on $SrTiO_3$ (001) have been solved by direct methods.[7, 8] Other reconstructions have been observed on pure $SrTiO_3$ (001), including the (1x1), (2x2), c(2x2), (4x4), c(4x4), (6x2), c(6x2), ($\sqrt{5}$x$\sqrt{5}$)R26°, and ($\sqrt{13}$x$\sqrt{13}$)R33.7°.[6, 9-19] Models have been proposed for many of these structures, though they are often inconsistent with one another, and theoretical models have also been developed, however these too remain contradictory.[18, 20-22]

One structure which has proven especially difficult to determine is the $SrTiO_3$(001)-c(6x2) surface reconstruction. The main challenge, as will be shown, is the fact that a single reconstruction is unable to adequately describe the surface, which probably is related to the high annealing temperature (1050-1100°C) required to form the surface. Instead, the equilibrium c(6x2) surface at the formation temperature is found to be short-range ordered, consisting of microscopic domains of four related structural motifs. Upon sufficiently rapid cooling, the surface structure is quenched, and the domains of the four motifs persist.

The c(6x2) has been previously reported by Jiang and Zegenhagen with scanning tunneling microscopy (STM) and low-energy electron diffraction[23, 24] and by Naito and Sato with reflection high-energy electron diffraction (RHEED).[17] The STM results are included here, and newly available x-ray diffraction results are also utilized. The c(6x2) studied by RHEED[17] was found to co-exist with domains of $(\sqrt{13}\times\sqrt{13})R33.7°$ and may likely be different from the surface studied here, as the surface preparation, which is known to play a large role, was different. As mentioned earlier, a (6x2) overlayer has also been observed on Nb doped $SrTiO_3$ (001),[19] however this structure is not the same as the c(6x2) reported here, since the (6x2) surface unit cell is not centered and thus has a different symmetry and structure.

Direct methods for surfaces based on diffraction data have been employed to solve numerous structures (for more information, see Refs[25, 26]), including two other surface reconstructions on $SrTiO_3$ (001), the (2x1) and c(4x2),[7, 8] as well as the $SrTiO_3$ (106) surface[27] and the $(\sqrt{5}\times\sqrt{5})R26.6°$ reconstruction on $LaAlO_3$ (001).[28] In some cases, e.g. the (2x1) and c(4x2) reconstructions on $SrTiO_3$, direct methods can be used to find all of the atoms in the surface structure. However, even with ideally perfect data, sometimes direct methods fail to resolve the atomic positions of certain atoms, particularly weakly scattering elements. Moreover, if disorder or twinning is present on the surface, structure completion (finding the full structure from an initial fragment) becomes exceedingly difficult. In this work on the c(6x2) reconstruction, direct methods alone did not result in a structure solution, but instead a combinatorial approach was taken that merged a variety of experimental and computational techniques and resulted in a model of the SrTiO$_3$(001)-c(6x2) surface that is consistent with all available experimental reports.

In more detail, the approach used in this work is to apply direct methods on a set of transmission electron and x-ray diffraction data[25, 26] in order to determine the approximate positions of the surface cations. Since the weak scattering of oxygen ions prevented conclusive determination of their positions from diffraction methods alone, computational combinatorial screening methods were used along with first-principles calculations to identify candidate oxygen configurations. First principles calculations were also used to more accurately determine the surface cation positions. These structural configurations were then used as input for structure refinement using surface x-ray data with the help of the Shelx-97[29] program, and simulated STM images from the output of the *ab initio* calculations were also compared with available experimental STM images as a final cross-check. The proposed surface structure for the c(6x2) reconstruction is consistent with all of the available experimental and computational evidence.

## 2. EXPERIMENTAL AND COMPUTATIONAL METHODS

Transmission electron diffraction (TED) experiments were conducted on samples prepared from single crystal, undoped SrTiO$_3$ (001) wafers (10x10x5mm$^3$, 99.95% pure). The wafers were cut into 3mm diameter discs using a rotary disc cutter, mechanically thinned to ~120μm, polished, dimpled, and ion milled to electron transparency with a 4.8 kV Ar$^+$ ion beam. Samples were annealed for 2-5 hours in a tube furnace at 1050 to 1100°C under a flow of high purity oxygen at atmospheric pressure in order to produce the reconstructed surface. Transmission electron microscopy images and off-zone diffraction patterns were obtained on the Hitachi ultra-high vacuum (UHV) H9000 electron microscope, operated at 300 kV. A series of off-zone diffraction patterns were recorded with exposure times ranging from 0.5 to 120 seconds. The negatives were scanned with a 25μm pixel size and digitized to 8 bits with an Optronics P-1000 microdensitometer. The diffraction intensities were then averaged with the *c2mm* Patterson plane group symmetry, yielding 58 independent intensities.

Surface x-ray diffraction (SXRD) experiments were performed at the BW2 wiggler beamline at the Hamburg Synchrotron radiation laboratory using radiation of 8 keV, monochromatized and sagittally focused by a pair of Si (111) crystals. Two single crystal SrTiO$_3$ (001) samples were annealed at the Max-Planck Institut in Stuttgart at 1100°C in flowing oxygen for about 2 hours. The samples were stored in an oxygen atmosphere container and shipped to another laboratory where they were characterized at room temperature by SXRD in air. One of the samples was measured in air a few days after the preparation. The second crystal was reloaded into a UHV chamber, exposed to a mild annealing in UHV at ~300°C and loaded into a small portable UHV chamber which was mounted on the diffractometer for the SXRD measurements. The acquisition of the diffraction data took approximately three days for each of the two samples. The stability of the surface over the acquisition period was ascertained by checking the stability of the (080) reflection at regular intervals, and integrated intensities were recorded for 263 in-plane reflections and 32 rods. The data were corrected for footprint and polarization, had reflections below the critical angle discarded, and were averaged using *C2mm* space group symmetry. The data taken for the two differently handled samples (oxygen annealed, oxygen & UHV annealed) were used separately for the structure refinement. See ref[30] for a copy of the SXRD data.

Scanning tunneling microscopy (STM) images were obtained using an Omicron "micro-STM" system operating under UHV conditions. The $SrTiO_3$ (001) c(6x2) sample, which was prepared outside the system by annealing at 1100°C in a flow of oxygen, was loaded into the UHV-STM system and annealed for approximately 10-15 minutes at 800°C in order to generate enough oxygen vacancies in the bulk to allow imaging by STM. Tungsten tips were used, and the STM scanner was calibrated with the use of the well-known Si(111)-(7x7) reconstruction. Images were obtained in constant current topography mode, and the sample was biased positively with respect to the tip, thus tunneling occurred into the empty states of the sample.

Direct methods were used to determine the scattering potential map of the surface structure based on the transmission electron and x-ray diffraction data. Direct methods solves the phase problem by utilizing probability relationships between the amplitude and phase of the diffracted beams. A set of phases is determined with the lowest figures of merit most consistent with scattering from discrete atoms and is combined with the measured beam amplitudes. By this approach, scattering potential maps and candidate structures can be generated from the diffraction data without the need for a structure guess.

First-principles (*ab initio*) density functional theory (DFT) calculations were performed using the Vienna *ab initio* simulation package (VASP),[31-34] which solves the DFT equations within the planewave-pseudopotential formalism. The SrTiO$_3$ (001) surface was represented by a surface slab model as illustrated in Fig. 1, with all atomic positions relaxed except for the center atomic layer which was held fixed at bulk positions and lattice parameters (determined in a separate bulk LDA calculation). The calculated lattice parameter (3.827 Å) is about 2% smaller than the experimental lattice parameter at room temperature (3.905 Å), which is typical for LDA calculations. Core-electrons were represented by Vanderbilt-type ultrasoft pseudopotentials[35, 36] (VASP library pseudopotentials "Ti", "Sr" and "O_s"), and electron exchange and correlation were treated in the local density approximation (LDA, Ceperley-Adler[37]). The planewave basis set was cut off at 270 eV. Simulated STM images were produced from the output of the *ab initio* calculations in the Tersoff-Hamann approximation,[38] which assumes that the point-like STM tip follows an isosurface of the local density of states within a specified energy window around the Fermi level. A relatively high isodensity surface lying very close to the surface was used, thus enabling us to use a smaller "vacuum" region in the supercell calculation. Simulated images were created using the integrated density of unoccupied states between 0 and +2.1 V relative to the Fermi level.

Surface x-ray diffraction data structure refinements were performed using the Shelx-97 code,[29] which is a widely used structural refinement program used in many fields including crystallography. The atomic positions for each of the plausible structures generated by DFT were input into the Shelx-97 program and refined primarily against the experimental data obtained in air. Since LDA calculations underestimate the lattice parameters, we scaled all atomic positions isotropically until the calculated lattice parameters matched the experimental value. This approach is preferable to imposing the experimental in-plane lattice parameters in the calculations, since the system would then contract perpendicular to the surface, resulting in an unphysical distortion that would be difficult to correct. The data were decomposed into 33 batches: 1 for the in-plane set and 32 for each of the rods, and each batch was given a different scale factor to account for experimental error in the data collection owing to changes in the sample-detector geometries upon measurements of different rods. Refinement parameters are given in the input (.ins) file and are described in the Shelx-97 manual. See ref[39] for a copy of the input (.ins) file and final $(F_c)^2$ values.

## 3. RESULTS

### A. Transmission electron microscopy

Dark field transmission electron microscopy images and off-zone diffraction patterns were obtained for the c(6x2) surface, as shown in Fig. 2. For the sample preparation techniques employed here, the c(6x2) surface reconstruction is highly reproducible and was found to be air-stable over a period of months. The dark field image in Fig. 2 shows a flat, faceted surface with large terraces separated by step bunches, and the c(6x2) surface reconstruction was found to cover the entire surface. Voids are also visible in the near-surface region of the sample, and similar morphologies have been observed for other reconstructed SrTiO$_3$ (001) surfaces.[40]

### B. Direct methods

Direct methods provided the scattering potential maps shown in Fig. 3 based on surface x-ray diffraction data [Fig. 3(a),(b),(c)] and transmission electron diffraction data [Fig. 3(d)]. Further analysis, based on symmetry and difference maps, indicated that the dark spots were titanium atom sites and that the surface contained no strontium atoms. Numerous attempts were made to refine a single structure with reasonable oxygen sites, but no single structure yielded good results. This occurred, as will be shown, because the surface is really a mixture of four different structural motifs. While the positions of the titanium atoms averaged over the four structural motifs could be determined in projection from the electron diffraction data and in three dimensions from the x-ray diffraction data, the positions of the surface oxygen atoms could not be determined owing to larger variation of the oxygen positions among the four motifs. This conclusion was reached by applying a combinational screening method in conjunction with first-principles methods to identify plausible oxygen configurations, with the averaged positions of the titanium atoms from the direct methods analysis used as the input for the screening method.

### C. Computational screening

The determination of the minimum energy oxygen configuration represented a challenging optimization problem, given the large configuration space that needed to be sampled and the presence of an enormous number of local minima in the system's potential energy surface, i.e. the energy of the system as a function of all atomic coordinates. Each local minimum is surrounded by a *basin of attraction*, where the set of all points connected to that local minimum follows a continuous path along which the energy decreases. The screening approach divided the optimization problem into (i) a discrete outer optimization problem over the different basins of attraction and (ii) a continuous inner optimization problem over atomic coordinates within each basin. The outer optimization problem scanned over basins and provided suitable starting points for the inner continuous optimization problem. In effect, each basin was represented by the selection of one starting point or *starting configuration* within it. The specific starting configuration in a basin was somewhat arbitrary since the inner continuous optimization problem should find the same local minimum regardless of the starting configuration used.

Starting configurations were constructed via enumeration of every possible placement combination of oxygen atoms on a lattice of plausible candidate sites. These candidate sites, shown in Fig. 4, are located at the midpoint of (1) every pair of titanium atoms separated by ≤ 4.25 Å and (2) every triplet of titanium atoms separated by ≤ 4.25 Å. Four-coordinated oxygen sites were not considered, because they either produced redundant sites or required at least one of the four titanium-oxygen bonds to be longer than 2.3 Å. One-fold coordinated oxygen sites on top of each of the four symmetrically distinct surface titanium atoms were considered as well.

The total number of possible ways to place oxygen atoms on the candidate sites was $2^{40}$, as there were 40 candidate oxygen sites in the asymmetric unit of the surface unit cell. All of these configurations possess, by construction, the *C2mm* space group determined from the experimental SXRD data. Since it would have been prohibitively computationally expensive to calculate the minimum energy within each basin associated with each of these starting configurations via first-principles methods, a hierarchy of increasingly precise criteria was utilized instead to screen out high-energy configurations.

At the coarsest level a simple geometric criteria was used, discarding configurations (1) with an oxygen deficiency exceeding two oxygen atoms per primitive surface unit cell, (2) with oxygen-oxygen bonds shorter than 1.8 Å, or (3) containing a titanium atom with a coordination number less than 3 or more than 6. These simple criteria reduced the number of plausible configurations to 17,095. While this number remained too large to be handled via *ab initio* methods, it was easily manageable using a simple electrostatic pair potential model, where the species Sr, Ti and O take the nominal charges 2+, 4+ and 2-, respectively, which could be used to efficiently identify the most promising configurations.

The electrostatic energy was calculated for each of the 17,095 candidate starting configurations previously identified. Note that the atomic positions were not relaxed in these calculations, otherwise the system would have collapsed to a point, since there were no short-range repulsive components in the interatomic forces. For nonstoichiometric structures (stoichiometry of $TiO_{2-x}$, where $x > 0$), the charges of all titanium atoms were reduced to 4 - 2$x$ to maintain charge balance, since the Fermi level would lie within the titanium bands under oxygen-deficient conditions.

At the end of this screening step, ~75 structures with the lowest electrostatic energy were retained, at each of the three surface stoichiometries considered (from zero to two oxygen vacancies per primitive surface unit cell). Fully relaxed LDA calculations were then performed for each of these ~75 structures using the VASP code. A representative structural geometry is illustrated in Fig. 1. Given the large number of candidate geometries, a relatively thin slab was used to represent the surface and the Brillouin zone was sampled at the $\Gamma$ point only, in order to limit the computational costs. After the structural relaxations, some of the starting configurations that were initially distinct actually converged to the same configuration: the ~75 starting configurations produced 64 distinct relaxed geometries. The convergence of some of the configurations towards each other was an indication that the initial partitioning of configuration space was sufficiently fine, as the "distance" between any two starting configurations was, on average, slightly smaller than the size of the typical basin of attraction.

The lowest energy configurations, i.e. structural motifs, thus identified for each of the three stoichiometries are shown in Fig. 5. These geometries were re-optimized using a thicker slab (twice the thickness shown in Fig. 1) and a finer k-point mesh (4×4×1) to yield more accurate energies. At $TiO_2$ stoichiometry, the lowest energy structure is labeled as "RumpledStoichiometric" [Fig. 5(a)]. The next lowest energy structure, labeled as "FlatStoichiometric" [Fig. 5(b)], is 0.37 eV less stable (per primitive unit cell).

At an oxygen content corresponding to one oxygen vacancy per primitive surface unit cell, the screening algorithm identified the "RumpledVacancy" as the lowest energy structure [Fig. 5(c)]. Visual inspection of that structure revealed that slight displacements along the surface normal of the titanium atoms near the center of the cell changed their coordination from 4-fold to 5-fold, resulting in another plausible structure, labeled "FlatVacancy" [Fig. 5(d)]. Given that the positions determined from direct methods were averaged over the four structural motifs, it is not entirely surprising that some adjustments would be required for the cation positions of any one of the individual structures. While the combinational screening did not find this structure automatically, as it assumed the positions of the Ti atom to be exact, it did identify a sufficiently similar structure to enable its discovery. During our initial lower-precision screening, the FlatVacancy structure appeared to have a lower energy than the RumpledVacancy structure. However, our more accurate re-optimization of the geometries revealed that the RumpledVacancy structure is the ground state at that composition, with an energy 0.26 eV/unit cell lower than the FlatVacancy structure.

At the composition corresponding to two oxygen vacancies per primitive surface unit cell, the "DoubleVacancy" structure was identified as the lowest energy structure [Fig. 5(e)]. The second most stable structure is more than 3 eV/unit cell less stable than the "DoubleVacancy" structure and can thus be ruled out.

The relative surface energy per primitive unit cell for each of these structural motifs were calculated and plotted as a function of oxygen chemical potential in FIG. 6. It is noted that the DoubleVacancy structure has a potential that is so high that its corresponding line lies far above the range of the figure and is therefore unlikely to be present on the surface. Since the exact surface energies are also a function of the Ti and Sr chemical potentials (which are difficult to infer from experimental conditions), we plot the surface energies relative to the RumpledStoichiometric surface energy. This difference in surface energies is sufficient to assess the relative stability of the motifs and offers the advantage that the contributions of the Ti and Sr chemical potentials cancel out exactly (because all motifs have the same number of Ti or Sr atoms). In contrast, the dependence on the O chemical potential cannot be similarly eliminated because the different motifs have different oxygen content.

The range of chemical potentials considered corresponds to temperatures ranging from 0K to 1300K. The oxygen chemical potential (in $O_2$ at atmospheric pressure) was obtained from the equation:

$$\mu_O(T) = (1/2)\mu_{O_2}(T) = (1/2) ( H_{LDA} + H(T) - H(0) - T\, S(T) )$$

where $H_{LDA} = -9.676$ eV (from a LDA calculation of an isolated $O_2$ molecule) and the following tabulated thermodynamic values from ref[41] were used: H(1300K) = 33344 J/mol, H(0K) = -8683 J/mol, , S(1300K) = 252.878 J/(mol K).

It is expected that the actual relative surface energies can be read off from FIG. 6 at a value of the oxygen chemical potential lying somewhere between the calculated extremes shown in the figure. At T = 0K the calculations have assumed zero entropy and therefore over-stabilize the stoichiometric phases, while at T = 1300K the calculations only account for the entropy of the gas phase and, since the free energy change of the solid phases may partially offset the $O_2$ chemical potential change, probably result an over-stabilization of the gas phase and of the nonstoichiometric phases.

The surface energy of the four structural motifs considered (RumpledStoichiometric, FlatStoichiometric, FlatVacancy, RumpledVacancy) lie within 0.4 eV/unit cell of each other for chemical potentials slightly below the 1300K value. The actual energy range is likely to be even smaller than our calculated range of 0.4 eV because our results neglect the contribution of lattice vibrations to the free energy. Structures that are very stable (low in energy) tend to be stiffer and therefore have a lower vibrational entropy and a more positive free energy. Conversely, vibrational effects tend to lower the free energies of high-energy structures, resulting in a reduction of the spread in the free energies. Thus, the surface energy differences lie in a range that is likely to be somewhat smaller than 0.4 eV, and thus comparable in magnitude to $k_BT$ at 1300 K (about 0.12 eV), making it quite plausible for the equilibrium surface structure to consist of a disordered mixture of these four structural motifs.

The accuracy of the approach we used to obtain the oxygen chemical potential is limited by the fact that the LDA tends to poorly predict the energy of an isolated molecule. An alternative approach, following ref,[20] that avoids calculating isolated molecule energies yielded qualitatively similar conclusions: All points where the different surface energies intersect lie between the values of the O chemical potential at 0K and 1300K.

The four structural motifs RumpledStoichiometric, FlatStoichiometric, FlatVacancy, and RumpledVacancy can be described using four atomic layers. Starting at the bottom for all motifs (in reference to the geometry shown in Fig. 5), there is a bulk-like TiO$_2$ layer followed above with a bulk-like SrO layer, and these two layers are nearly identical in all four motifs. The next TiO$_2$ layer up is similar in all structural motifs and has a rumpled bulk-like structure, with relaxations along the direction normal to the surface of at most ~0.12*$a_{bulk}$. Finally the topmost layer is different for each of the four motifs in the number and placement of the oxygen atoms: the top layer has a Ti$_{20}$O$_{40}$ stoichiometry in the stoichiometric structure centered unit cell and has a Ti$_{20}$O$_{38}$ stoichiometry in the vacancy structure centered unit cell. Note that the titanium positions are nearly identical in all structures. See ref[42] for the atomic positions of the four structural motifs.

For each structural motif, the topmost layer contains a zig-zag along the *b* (short axis) direction of five-fold co-ordinated titanium atoms in the form of truncated octahedra. In the centered unit cell, the two zig-zags are located at approximately ¼ and ¾ along the length of the long (*a*) axis (see Fig. 5), and the relative orientation of the truncated octahedra along the zig-zags is the same for three of the four structures and is reversed in the RumpledVacancy structure. In the rumpled structures (RumpledStoichiometric and RumpledVacancy) the zig-zag is elevated normal to the surface relative to the center of the unit cell, and in the flat structures (FlatStoichiometric and FlatVacancy) the center of the unit cell is at approximately the same elevation as the zig-zag. Accordingly, the titanium atoms at the center of the unit cell (not part of the zig-zag) in the rumpled structures are coordinated to the bulk-like layer below, while in the flat structures they are not. The coordination of the titanium atoms at the center of the cell is the driving force for the placement of the singly-coordinated oxygen (if any) in the various structures. In the structures containing a singly coordinated oxygen, i.e. the RumpledStoichiometric, FlatStoichiometric, and RumpledVacancy structures, the singly coordinated Ti-O bonds are 1.68Å, 1.65Å, and 1.65 Å long, respectively, indicating double bond (titanyl) character. Essentially, the differences among the four structures lie in the relative orientation of the truncated octahedra in the zig-zag chain, the elevation and coordination of the titanium atoms located in the center of the unit cell, and the placement of the singly-coordinated oxygen (if any) at the surface.

### D. STM experiment and simulation

In STM images taken under empty-state bias conditions (Fig. 7), the c(6x2) reconstruction appears as bright rows with a spacing of 11.7 Å (cf. with 11.715 Å for ½ the c(6x2) long axis length, 23.43 Å). Confirmed to be c(6x2) by low energy electron diffraction, the reconstruction was found to cover the surface uniformly wherever probed by the STM. In large-scale images (not shown), the rows appear to be aligned with equal probability along the [100] or [010] crystal directions, and in addition to the rows, bright protrusions situated on the rows can be seen randomly distributed over the surface with a density of approximately one for every three c(6x2) centered unit cells. It is noted that sufficient conductivity in $SrTiO_3$ is achieved with an overall carrier density due to oxygen vacancies smaller than $10^{18}$ $e$/cm$^3$, i.e., roughly 1 out of every 30 neighboring oxygen atoms missing. It is expected that the density of oxygen vacancies on the surface may be slightly higher, but still low compared to the density of observed contrast variations. Furthermore, preliminary experimental studies in which SXRD data were collected on samples used for STM and LEED have evidenced that the UHV anneal prior to STM measurements has a minimal effect on the c(6x2) structure.

The simulated STM images, shown in Fig. 5 for each of the structural motifs considered, confirm that in empty-state only titanium atoms image brightly, while oxygen atoms are dark, and thus the experimentally observed rows are in fact the zig-zags of truncated octahedra discussed earlier. Note that the point-like tip approximation and the tracing of a relatively high isodensity surface resulted in simulated STM images of higher resolution (sharper) than the experimental image. Upon detailed investigation of the experimental image, changes in the relative orientation of the zig-zags can be seen occasionally from one row to another, evidence of domain boundaries between different structural motifs.

Upon inspection of the simulated STM images from the structural motifs alone, the bright protrusions observed in the experimental STM images are not accounted for. Based on the previous observation that the STM is imaging titanium atoms, it was determined that the contrast of the bright protrusion is due to excess non-periodic titanium atoms along the zig-zag. Upon studying plausible structures, a likely location for the titanium atom is readily apparent in the RumpledStoichiometric structure. This plausible geometry is suggested by the fact that the two singly-coordinated oxygen atoms are at just the right position so that an additional $TiO_2$ unit could be placed on the surface, and the inserted titanium atom would have a 4-fold coordination and the inserted oxygen atoms would complete the octahedral coordination of the truncated octahedra in the zig-zag. To clarify the nature of these bright protrusions, a simulated STM image was generated of the RumpledStoichiometric surface with an additional $TiO_2$ unit located on the zig-zag [see Fig. 5(f)], and the calculated STM image of this surface is in qualitative agreement with the experimentally observed bright protrusions. Note that the final surface stoichiometry is $Ti_{21}O_{42}$ for one unit added per centered unit cell, and thus $TiO_2$ is added to the structure in a stoichiometric manner. See ref[43] for the DFT refined positions of the $TiO_2$ unit.

### E. **Structure refinement**

To substantiate the proposed c(6x2) surface structure model, refinement with XRD data was carried out by means of the Shelx-97 refinement program.[29] Use of this program allowed for the refinement of the complicated, multi-domained c(6x2) structure through partial occupancies of atom sites. Figures of merit including weighted R-values ($wR^2$) and Goodness of Fit were employed as a gauge for the quality of the refinement, and the Hamilton R-factor ratio[44] was utilized to compare $wR^2$ values for structural refinements with various numbers of parameters. The absolute values of the figures of merit do not hold much meaning outside of this study, as this is not a standard Shelx structural refinement, but rather the figures of merit are used to compare models relative to one another. Further, it is important to note that one should not expect a perfect fit between the DFT-calculated positions and the refined positions. Both methods invoke approximations: notably, the refinement process relies on partial occupancies to model disorder, and the DFT calculations neglect thermal expansion, which could affect the average positions of atoms in low-symmetry environments and have an accuracy limited by the unavoidable approximation of the exchange-correlation functional and, to a lesser extent, by the finite *k*-point mesh and energy cutoff.

The four DFT-relaxed structural motifs were refined independently for 25 least squared cycles, and the structures had three bulk-like layers below the surface atoms, as illustrated in the cartoons of Fig. 5. Additionally, in order to better represent the surface from which the data were acquired, all four structural motifs were combined and refined simultaneously for 25 least squared cycles. In this case, the combined structure had the same three bulk-like layers as the other structures but had a surface containing the atoms from all four structural motifs. The occupancies for the surface atoms representing the four motifs FlatStoichiometric, FlatVacancy, RumpledStoichiometric, RumpledVacancy ($x_{FS}$, $x_{FV}$, $x_{RS}$, $x_{RV}$, respectively) were constrained such that the sum of the four occupancies summed to 1, and initially each motif was assigned an occupancy of 25%.

A $TiO_2$ unit was placed on top of the surface's zig-zag with occupancy $x_{TiO2}$ to correlate with the bright protrusions in the experimental STM images. Owing to the symmetry constraints of the refinement, the $TiO_2$ was added in a periodic fashion, because adding a single $TiO_2$ unit in the unit cell would require a reduction in the symmetry, therefore increasing the number of parameters (*p*), which is undesirable. Thus to model the non-periodic nature of the $TiO_2$ unit, the occupancy ($x_{TiO2}$) was allowed to vary as an independent variable.

Table I shows the figures of merit for each of the structural refinements: four motifs combined plus the $TiO_2$ unit, four motifs combined without $TiO_2$ unit, RumpledStoichiometric, FlatStoichiometric, FlatVacancy, and RumpledVacancy. It is important to note that the positions relaxed by the DFT calculations did not change much upon refinement, providing strong evidence that they are appropriate models. Using the Hamilton R-factor ratio,[44] the structure with the four motifs combined fits the data better than any of the other individual models with greater than 90% certainty. Other models were tested, including structures composed of combinations of two or three of the structural motifs and structures incorporating the DoubleVacancy motif, however these refinements tended to be inferior and supported the four structural motif model.

The figures of merit for the individual structure refinements are similar for the FlatVacancy, RumpledStoichiometric and FlatStoichiometric structures and showed a worse fit for the RumpledVacancy structure, all in qualitative agreement with the relative surface energy values. For the four motifs combined structure, the final values for $x_{FS}$, $x_{FV}$, $x_{RS}$, $x_{RV}$ each remained close to 25%, i.e. each structural motif is present on approximately ¼ of the surface. The $TiO_2$ unit ($x_{TiO2}$) is situated on roughly 15 to 45% of the c(6x2) surface unit cells, which agrees well with the experimental STM measurement of approximately 33%. Data from the second sample, also annealed in $O_2$ at 1100°C but subsequently annealed in UHV at 300°C, also gave similar occupancies for $x_{FS}$, $x_{FV}$, $x_{RS}$, $x_{RV}$ and $x_{TiO2}$ in the four motifs combined structure, which is expected since the oxygen chemical potential at 1000°C in $O_2$ and at 300°C in UHV are similar (-3.2eV and –2.57eV, respectively).

## 4. DISCUSSION

A model for the structure of the c(6x2) reconstruction has been proposed, and unlike the (2x1) and c(4x2) reconstructions on SrTiO$_3$(001), the c(6x2) structure solution was not explicitly provided from direct methods analysis alone. Of the three reconstructions, the c(6x2) forms at the highest temperature, 1050-1100°C, compared to 850-930°C for c(4x2) and 950-1050°C for (2x1), and is therefore, not surprisingly, the most complex structure. The surface is composed of short-ranged ordered domains of four related structures, ranging from stoichiometric to slightly reduced (one oxygen vacancy per primitive surface unit cell), each present on approximately ¼ of the total surface area. At the temperature and oxygen partial pressure required for the formation of the c(6x2) surface reconstruction, the formation energies for these structures are quite comparable, and the surface thus takes the form of a random (although short-range-ordered) mixture of these four structural motifs. A rough approximation for the entropy of mixing is 1.39*kT per unit cell area, which at 1100°C is 0.164eV. This value represents the upper bound, as it neglects domain boundary energy and assumes the structure of one unit cell does not influence the structure of neighboring cells. Additionally the TiO$_2$ unit, which is present non-periodically on the surface, also results in an entropic free energy gain for the surface.

The proposed c(6x2) structure, while the most complicated reconstruction on SrTiO$_3$, shows similarities to the (2x1) and c(4x2) structures.[7, 8] All three structures are terminated with a Ti$_y$O$_x$ surface layer – that is, there are no strontium atoms on the surface. The c(4x2) and (2x1) reconstructions are composed of a single $TiO_2$-stoichiometry overlayer above bulk-like $TiO_2$, and the difference between the c(4x2) and (2x1) structures is the distribution of the surface Ti among the possible sites. The c(6x2), on the other hand, has a thicker (more than one) $TiO_x$ overlayer above the bulk-like $TiO_2$ layer. Furthermore, the c(4x2) and (2x1) structures have titanium cations present on the surface solely in the form of 5-fold, truncated octahedra, and while the c(6x2) reconstruction does have titanium cations in 5-fold truncated octahedra, titanium cations are also present in the surface structure with 4-fold coordination. The most striking difference is the fact that the c(6x2) reconstruction is composed of multiple related, but different, structural domains, while the c(4x2) and (2x1) reconstructions are single-structure surfaces. Finally, $TiO_2$ units are stabilized on the surface of (001)-$SrTiO_3$ c(6x2), but no evidence exists for this type of behavior on the c(4x2) or (2x1) surfaces.

It is believed that the c(6x2) surface is likely to be the most catalytically active surface of (001) $SrTiO_3$. With titanium atoms present in multiple coordination geometries and oxidation states, the surface would likely be able to bind reactant molecules and promote redox-type reactions. The c(6x2) reconstruction (as well as the (2x1)) contains Ti=O (titanyl) groups which have recently been implicated with catalytic activity on the (011) surface of rutile $TiO_2$.[45] Furthermore, the presence of the $TiO_2$ unit suggests the ability of the surface to stabilize reaction intermediates, and research is currently underway to investigate the adsorption, desorption, and reactivity of methyl radicals on the various reconstructions of $SrTiO_3$ (001).

## 5. CONCLUSIONS

In conclusion, a model for the atomic scale structure of the $SrTiO_3$(001)-c(6x2) surface reconstruction has been proposed. The surface reconstruction is formed at high temperatures (1050-1100$^o$C) in oxidizing conditions and is highly stable and reproducible. The surface is composed of domains of similar but distinct structures, and additionally, $TiO_2$ units are randomly distributed on the surface. While the structure solution method was not conventional, we have acquired the maximum amount of information through a combination of techniques. Transmission electron diffraction and surface x-ray diffraction provided the positions of the surface titanium atoms averaged over the four structural motifs, and the *ab initio* screening technique proved to be indispensable for the determination of oxygen positions, as well as the titanium positions along the z-direction. Adaptation of the Shelx-97 program for structure refinement against surface x-ray data merged theory with experiment to corroborate the model, and finally STM simulations confirmed consistency with experimental observations.

## ACKNOWLEDGEMENTS

CL, KRP, and LDM were supported by the Chemical Sciences, Geosciences and Biosciences Division, Office of Basic Energy Sciences, U.S. Department of Energy Office of Science Grant #DE-FG02-03ER15457, and NE and OW were supported by the EMSI program of the National Science Foundation and the U.S. Department of Energy Office of Science Grant #CHE-9810378, all at the Northwestern University Institute for

Environmental Catalysis. AvdW and MA were supported by the National Science Foundation under program NSF-MRSEC DMR-00706097, and through TeraGrid computing resources provided by NCSA and SDSC. EL was supported by the National Science Foundation via grant #DMR-9214505. AK and JZ were supported by the German BMBF under contracts 05SE8GUA5 and 05KS1GUC3.

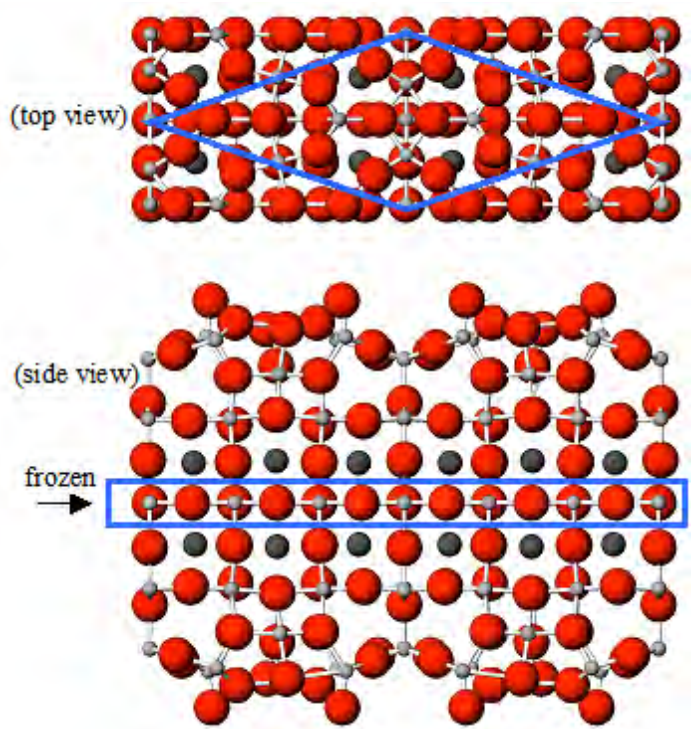

**FIG. 1.** (Color online) Geometry employed in the *ab initio* calculations, with the primitive c(6x2) surface unit cell outlined (representative structure shown). Large red spheres are oxygen, small light gray spheres are titanium, and medium dark gray spheres are strontium. The geometries of the two lowest energy structures at each composition were also re-optimized using a thicker slab (including 4 strontium layers instead of 2) in which the middle layer (containing Ti and O) was kept frozen.

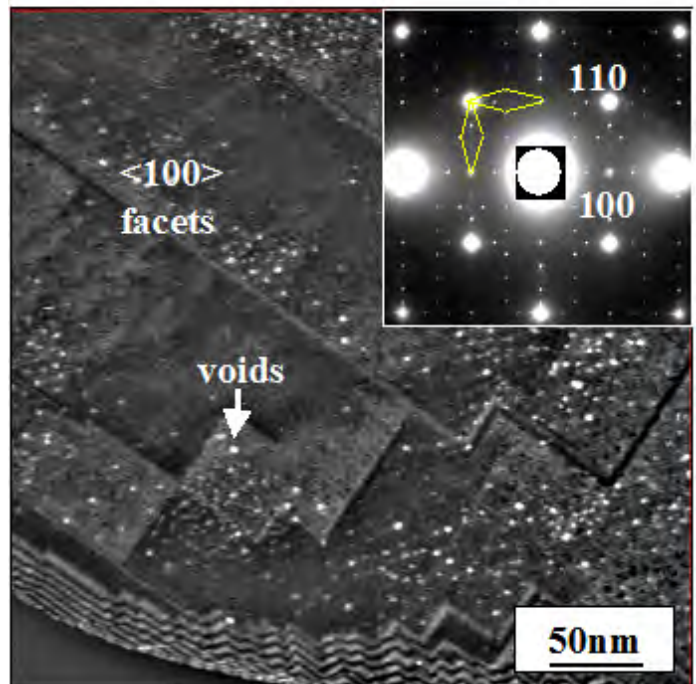

**FIG. 2.** Dark field image and transmission electron diffraction data (inset) from the c(6x2) surface. Primitive reciprocal unit cells for the two surface domains unit cell are outlined. Adapted from ref[40].

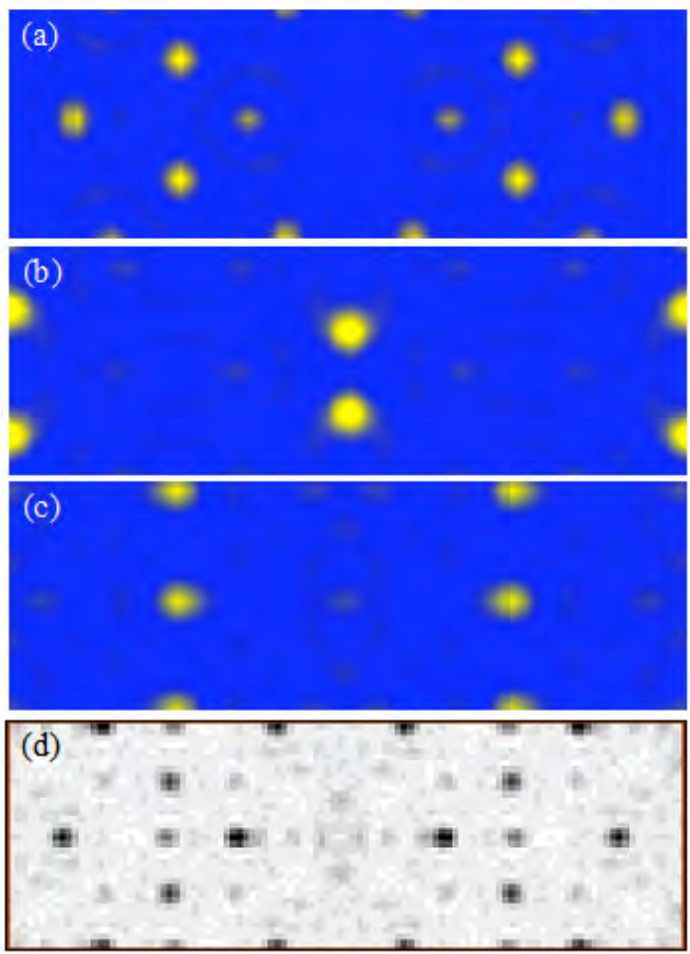

**FIG. 3.** (a), (b), (c) (Color online) Electron density maps for the centered c(6x2) unit cell from SXRD direct methods at $z = 3.6$ Å, $z = 2.8$ Å, and $z = 2.0$ Å above the first bulk-like $TiO_2$ layer, respectively. Regions of high electron density (possible atomic sites) are yellow/light. (d) Scattering potential map (projected) for the centered c(6x2) unit cell from TED direct methods. Regions of high scattering potential (possible atomic sites) are black.

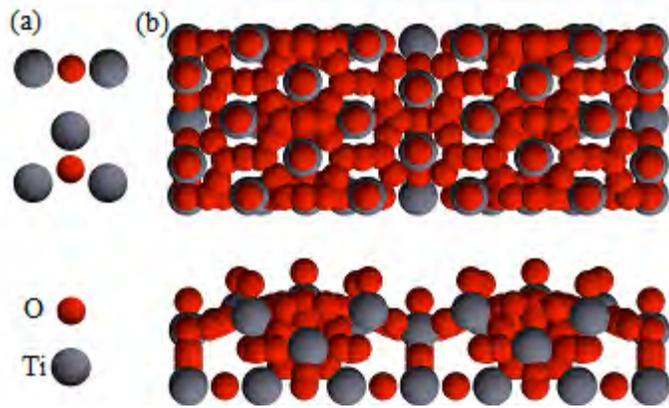

**FIG. 4.** (Color online) (a) Geometric rules used to generate candidate O sites shown in (b). Top panel is view towards the surface while the bottom panel is a side view with the free surface pointing upward.

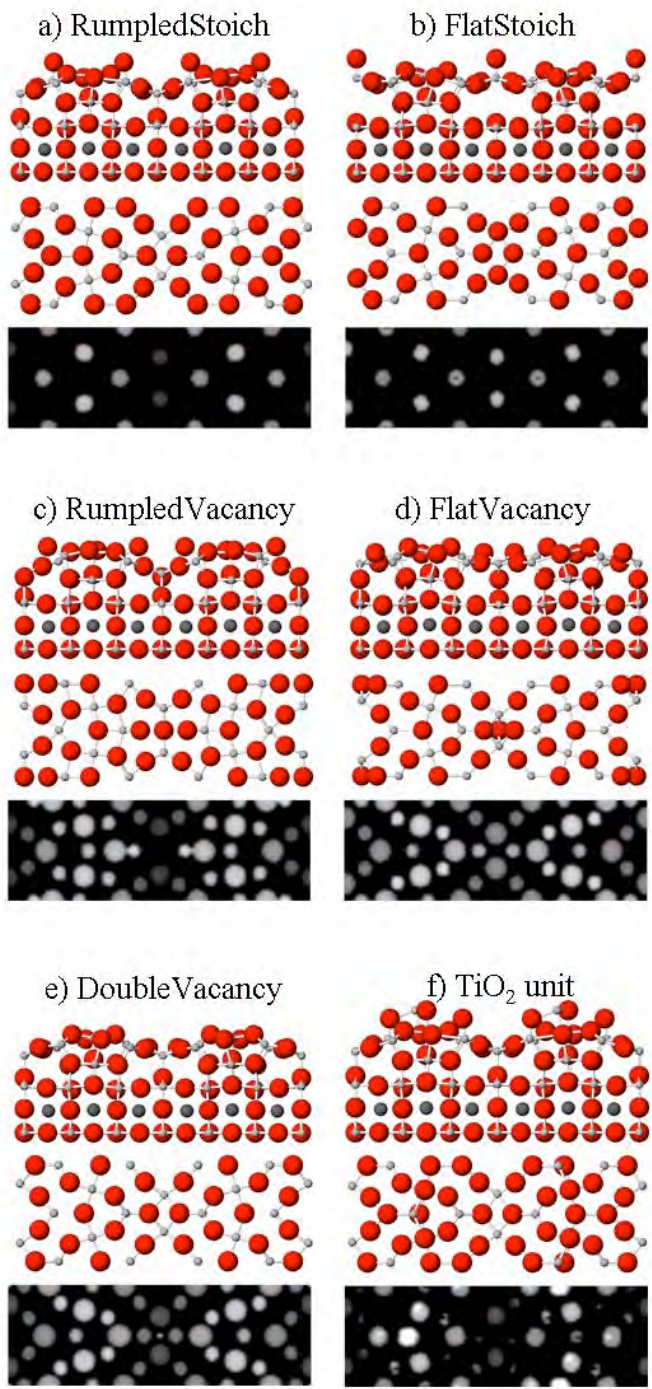

**FIG. 5.** (Color online) Candidate surface reconstructions showing side view, top view (showing only atoms in the topmost surface layer), and simulated STM image. Large red spheres are oxygen, small light gray spheres are titanium, and medium dark gray spheres are strontium.

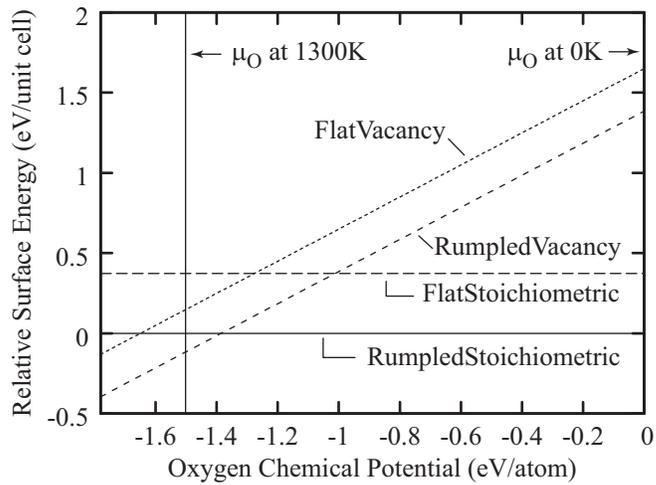

**FIG. 6.** Relative surface energy per primitive surface unit cell of the four proposed surface motifs as a function of oxygen chemical potential. The surface energies are given relative to the RumpledStoichiometric structure and the chemical potential is relative to its value at 0K.

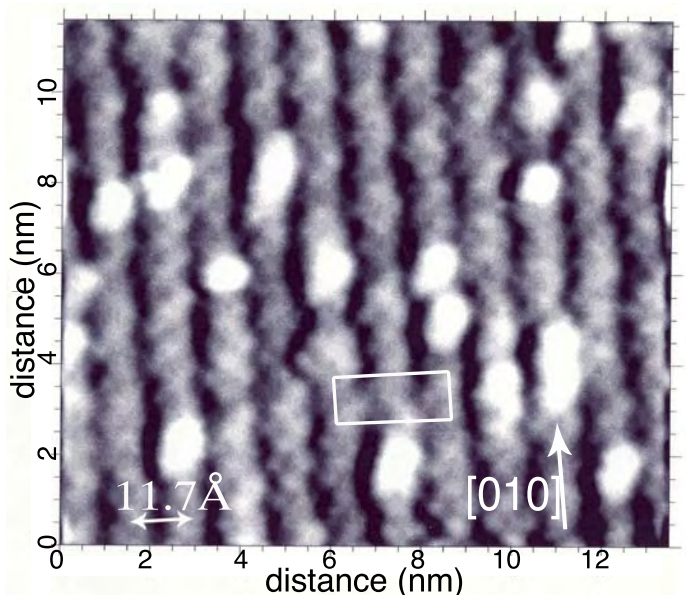

**FIG. 7.** High resolution STM image of the c(6x2) surface reconstruction ($V_s = 2.1V$, $I = 0.28nA$). The c(6x2) centered unit cell is outlined. Adapted from ref[23].

| Model | # LS | # data (n) | # parameters (m) | $wR^2$ | Goodness of Fit |
|---|---|---|---|---|---|
| Four motifs combined, with $TiO_2$ unit | 25 | 848 | 286 | 0.65 | 5.67 |
| Four motifs combined, without $TiO_2$ unit | 25 | 848 | 280 | 0.65 | 5.65 |
| RumpledStoichiometric only | 25 | 848 | 158 | 0.74 | 6.27 |
| FlatStoichiometric only | 25 | 848 | 158 | 0.74 | 6.33 |
| RumpledVacancy only | 25 | 848 | 157 | 0.77 | 6.69 |
| FlatVacancy only | 25 | 848 | 157 | 0.73 | 6.25 |

**TABLE I:** Figures of merit for refinement of DFT-relaxed structures against SXRD data.

# References

[1] V. E. Henrich and P. A. Cox, *The Surface Science of Metal Oxides* (Cambridge University Press, 1994).

[2] Materials Research Symposium Proceedings 401: Epitaxial Oxide Thin Films II, Pittsburgh PA, edited by J. S. Speck, D. K. Fork, R. M. Wolf, T. Shiosaki (1996).

[3] R. A. McKee, F. J. Walker, and M. F. Chisholm, Physical Review Letters **81**, 3014 (1998).

[4] R. A. McKee, F. J. Walker, and M. F. Chisholm, Science **293**, 468 (2001).

[5] J. G. Mavroides, J. A. Kafalas, and D. F. Kolesar, Applied Physics Letters **28**, 241 (1976).

[6] B. Cord and R. Courths, Surface Science **162**, 34 (1985).

[7] N. Erdman, K. R. Poeppelmeier, M. Asta, O. Warschkow, D. E. Ellis, and L. D. Marks, Nature **419**, 55 (2002).

[8] N. Erdman, O. Warschkow, M. Asta, K. R. Poeppelmeier, D. E. Ellis, and L. D. Marks, Journal of the American Chemical Society **125**, 10050 (2003).

[9] T. Nishimura, A. Ikeda, H. Namba, T. Morishita, and Y. Kido, Surface Science **421**, 273 (1999).

[10] V. Vonk, S. Konings, G. J. van Hummel, S. Harkema, and H. Graafsma, Surface Science **595**, 183 (2005).

[11] Q. D. Jiang and J. Zegenhagen, Surface Science **338**, L882 (1995).

[12] T. Matsumoto, H. Tanaka, T. Kawai, and S. Kawai, Surface Science **278**, L153 (1992).

[13] P. J. Moller, S. A. Komolov, and E. F. Lazneva, Surface Science **425**, 15 (1999).

[14] H. Tanaka, T. Matsumoto, T. Kawai, and S. Kawai, Japanese Journal of Applied Physics Part 1-Regular Papers Short Notes & Review Papers **32**, 1405 (1993).

[15] T. Kubo and H. Nozoye, Physical Review Letters **86**, 1801 (2001).

[16] M. S. M. Gonzalez, M. H. Aguirre, E. Moran, M. A. Alario-Franco, V. Perez-Dieste, J. Avila, and M. C. Asensio, Solid State Sciences **2**, 519 (2000).

[17] M. Naito and H. Sato, Physica C **229**, 1 (1994).

[18] T. Kubo and H. Nozoye, Surface Science **542**, 177 (2003).

[19] M. R. Castell, Surface Science **516**, 33 (2002).

[20] K. Johnston, M. R. Castell, A. T. Paxton, and M. W. Finnis, Physical Review B **70**, 085415 (2004).

[21] O. Warschkow, M. Asta, N. Erdman, K. R. Poeppelmeier, D. E. Ellis, and L. D. Marks, Surface Science **573**, 446 (2004).

[22] L. M. Liborio, C. G. Sanchez, A. T. Paxton, and M. W. Finnis, Journal of Physics-Condensed Matter **17**, L223 (2005).

[23] Q. D. Jiang and J. Zegenhagen, Surface Science **367**, L42 (1996).

[24] Q. D. Jiang and J. Zegenhagen, Surface Science **425**, 343 (1999).

[25] L. D. Marks, E. Bengu, C. Collazo-Davila, D. Grozea, E. Landree, C. Leslie, and W. Sinkler, Surface Review and Letters **5**, 1087 (1998).

[26] L. D. Marks, N. Erdman, and A. Subramanian, Journal of Physics-Condensed Matter **13**, 10677 (2001).

[27] X. Torrelles, J. Zegenhagen, J. Rius, T. Gloege, L. X. Cao, and W. Moritz, Surface Science **589**, 184 (2005).

[28] C. H. Lanier, J. M. Rondinelli, B. Deng, R. Kilaas, K. R. Poeppelmeier, and L. D. Marks, Physical Review Letters **98** (2007).

[29] G. M. Sheldrick, (1993-1997).

[30] EPAPS Document No. 1: copy of SXRD data.

[31] G. Kresse and J. Hafner, Physical Review B **47**, 558 (1993).

[32] G. Kresse and J. Hafner, Physical Review B **49**, 14251 (1994).

[33] G. Kresse and J. Furthmuller, Computational Materials Science **6**, 15 (1996).

[34] G. Kresse and J. Furthmuller, Physical Review B **54**, 11169 (1996).

[35] D. Vanderbilt, Physical Review B **41**, 7892 (1990).

[36] G. Kresse and J. Hafner, Journal of Physics-Condensed Matter **6**, 8245 (1994).

[37] D. M. Ceperley and B. J. Alder, Physical Review Letters **45**, 566 (1980).

[38] J. Tersoff and D. R. Hamann, Physical Review B **31**, 805 (1985).

[39] EPAPS Document No. 2: copy of input (.ins) file and final $(F_c)^2$ values.

[40] N. Erdman and L. D. Marks, Surface Science **526**, 107 (2003).

[41] M. W. Chase, C. A. Davies, J. R. Downey, D. J. Frurip, R. A. McDonald, and A. N. Syverud, Journal of Physical and Chemical Reference Data **14**, 927 (1985).

[42] EPAPS Document No. 3: atomic positions of the four structural motifs.

[43] EPAPS Document No. 4: DFT refined positions for $TiO_2$ unit.

[44] W. C. Hamilton, Acta Crystallographica **18**, 502 (1965).

[45] T. J. Beck, A. Klust, M. Batzill, U. Diebold, C. Di Valentin, and A. Selloni, Physical Review Letters **93**, 6104 (2004).